\documentstyle[seceq,twocolumn,epsf]{jpsj}

\def\H{{\cal H}}
\def\HA{{\cal H}_{\rm A}}
\def\HB{{\cal H}_{\rm B}}
\def\HC{{\cal H}_{\rm C}} 
\def\JA{J_{\rm A}}
\def\JB{J_{\rm B}}
\def\JC{J_{\rm C}}

\def\mpynn{$m$-MPYNN$\cdot$BF$_4$}

\def\v#1{\mib #1}

\def\runtitle{Uniform and Distorted Kagom\'e Heisenberg Antiferromagnets
}

\def\runauthor{Kazuo {\sc Hida}}

\title
{
Magnetization Process of the $S=1$ and $1/2$ Uniform and Distorted Kagom\'e Heisenberg Antiferromagnets
}

\author
{Kazuo {\sc Hida}
\footnote{e-mail: hida@phy.saitama-u.ac.jp}}

\inst
{
Department of Physics, Faculty of Science,\\ Saitama University, Saitama 338-8570
}

\recdate
{
\today
}

\abst
{
The magnetization process of the $S=1$ and $1/2$ kagom\'e Heisenberg antiferromagnet is studied by means of the numerical exact diagonalization method. It is found that the magnetization curve at zero temperature has a plateau at $1/3$ of the full magnetization. In the presence of $\sqrt{3} \times \sqrt{3}$ lattice distortion, this plateau is enhanced and eventually the ferrimagnetic state is realized. There also appear the minor plateaux above the main plateau. The physical origin of these phenomena is discussed.  }
\kword
{
kagom\'e Heisenberg antiferromagnet, numerical diagonalization, magnetization plateau
}

\begin{document}
\maketitle

\section{Introduction}

The kagom\'e Heisenberg antiferromagnet (KHAF) has been extensively studied theoretically and experimentally because of the interest in the interplay of the strong quantum fluctuation and the highly frustrated nature of the lattice structure. So far, most of the attempts have been focused on the ground state and low lying excitations of the uniform KHAF. In both $S=1/2$\cite{ze1,ey1,nm1,wal1}  and $S=1$\cite{kh1} cases, it is known that the ground state is a spin singlet state and the magnetic excitation has a finite energy gap. In the $S=1/2$ case, there are a number of singlet excitations below the first triplet excitation possiblly down to zero energy in the thermodynamic limit.\cite{wal1} On the other hand, the singlet excitations also have finite energy gaps in the $S=1$ case. The present author proposed the hexagonal singlet solid (HSS) picture for the ground state of $S=1$ KHAF.\cite{kh1} 

In the present work, we first investigate the magnetization process of this model. As examples of the frustrated quantum magnets, the magnetization process of the triangular Heisenberg magnet is widely studied.\cite{collins,honecker,miya} The magnetization plateau at one-third of the full magnetization is well estabilshed in the Ising-like classical Heisenberg model as a typical example of the 'classical' magnetization plateau.\cite{miya} Although this plateau can be well understood in terms of the classical spin configuration, it has been observed even in the isotropic Heisenberg model with $S=1/2$ which has strong quantum fluctuation.\cite{honecker} On the other hand, the honeycomb lattice $S=1/2$ Heisenberg antiferromagnet has no plateau.\cite{honecker} In this context, it is interesting to investigate the case of kagom\'e lattice which has the triangles and hexagons at the same time.

As a real material, Wada and coworkers\cite{wada1,awaga1,wata1}  have investigated the magnetic behavior of \mpynn which can be regarded as the $S=1$ kagom\'e antiferromagnet. This material, however, undergoes a structual transformation at 128.7K to the distorted phase with $\sqrt{3}\times\sqrt{3}$ structure.\cite{kambe} Motivated by this observation, we further investigate the effect of such lattice distortion to the ground state and magnetization process in both $S=1/2$ and $S=1$ cases. We find that the plateau is enhanced by the lattice distortion and eventually the ferrimagnetic ground state is realized. 

This paper is organized as follows: In \S 2, the model Hamiltonian is presented. The numerical results for the magnetization curves are presented in \S 3. The physical pictures of the plateau state and ferrimagnetic state are also explained. The last section is devoted to a summary and discussion.

\section{Model Hamiltonian}

The Hamiltonian of our model is given by, 
\begin{equation}
\label{ham1}
\H = J \sum_{<i,j>} \v{S}_{i} \v{S}_{j}-g\mu_{\scriptsize\rm B}H\sum_{i=1}^{N}S_i^z,
\end{equation} 
where $H$ is the magnetic field, $g$ the electronic $g$-factor, $\mu_{\scriptsize\rm B}$ the Bohr magneton and $\v{S}_{i}$ is the spin operator. In the following, we take the unit $g\mu_{\scriptsize\rm B}=1$. The summation $\sum_{<i,j>}$ is taken over all the nearest neighbour pairs of sites of the kagom\'e  lattice. 

\begin{figure}
\epsfxsize=70mm 
\centerline{\epsfbox{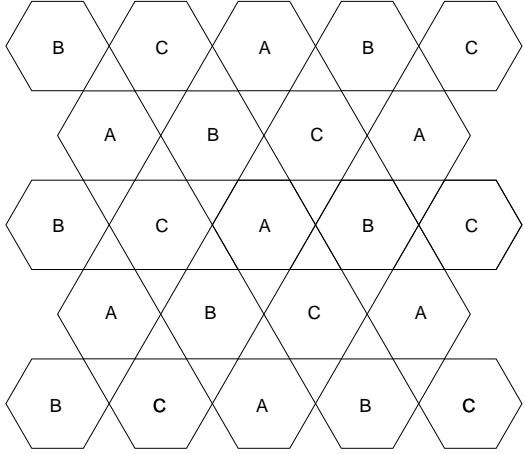}}
\caption{$\sqrt{3}\times\sqrt{3}$ distorted kagom\'e lattice. }
\label{fig2}
\end{figure}

 We also consider the $\sqrt{3}\times\sqrt{3}$ distorted version of this model given by,
 
\begin{eqnarray}
\label{ham2}
\H &=& \HA + \HB+ \HC-g\mu_{\scriptsize\rm B}H\sum_{i=1}^{N}S_i^z, \\
\H_{\alpha} &=& J_{\alpha}\sum_{<i,j> \in \alpha}\v{S}_{i} \v{S}_{j}, 
\end{eqnarray} 
where $\sum_{<i,j> \in \alpha}$  represent the summation over the bonds around the type-$\alpha$ $ (\alpha = \mbox{A, B} \ \mbox{or  C})$ hexagons, which are depicted in Fig. \ref{fig2}.  
\begin{figure}
\epsfxsize=80mm 
\centerline{\epsfbox{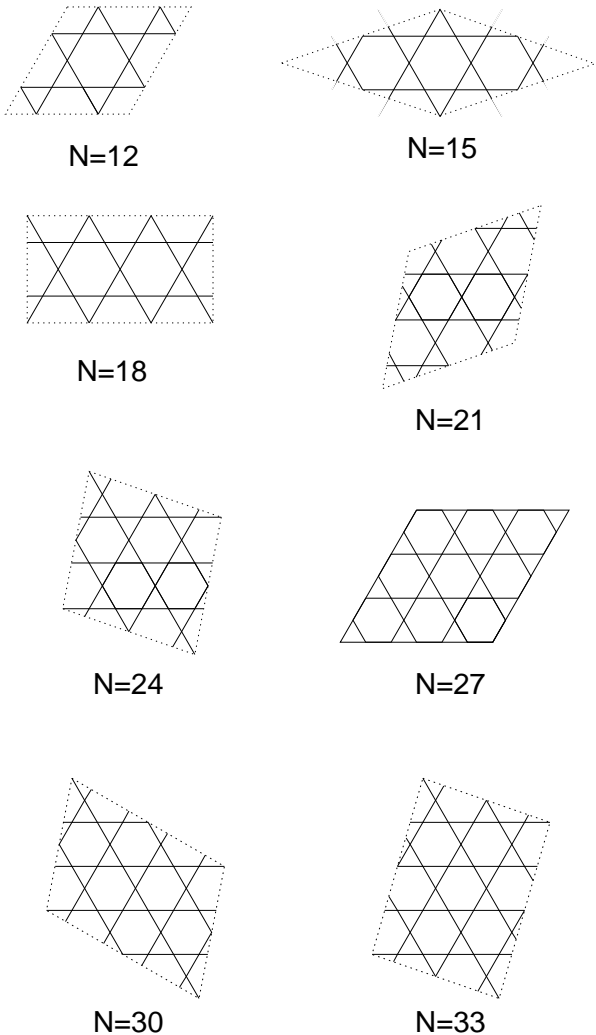}}
\caption{Clusters used for numerical diagonalization. }
\label{clusters}
\end{figure}

\section{Magnetization Curves}

The Hamiltonians of the finite size clusters shown in Fig. \ref{clusters} are diagonalized numerically to obtain the zero temperature magnetization curves. The periodic boundary condition is assumed.

\subsection{Uniform case}

\begin{figure}

\centerline{\epsfxsize=80mm\epsfbox{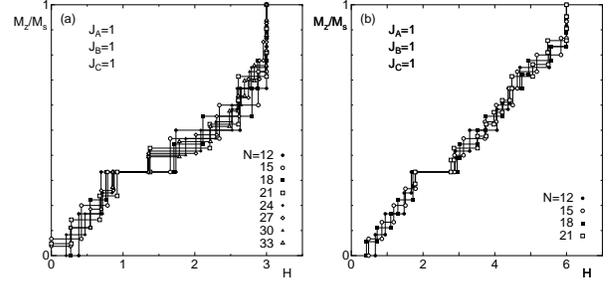}}
\caption{The magnetization curves of (a) $S=1/2$ ($N = 12,15,18,21,24,27,30$ and 33) and (b) $S=1$ KHAF ($N = 12,15,18$ and 21). Here and in the following figures the magnetic field is normalized by $J/g\mu_{\scriptsize\rm B}$.}
\label{uni}
\end{figure}

\begin{figure}

\centerline{\epsfxsize=80mm\epsfbox{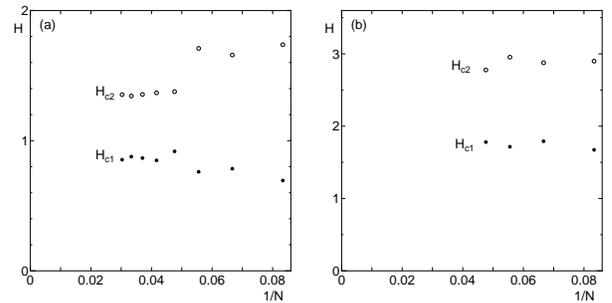}}
\caption{The system size dependence of the position of the plateau of (a) $S=1/2$ and (b) $S=1$ KHAF. The lower (upper) end of the plateau is denoted by $H_{\scriptsize\rm c1} (H_{\scriptsize\rm c2})$.}
\label{pla}
\end{figure}

For $S=1/2$, we use the clusters with $N=12$, 15, 18, 21, 24 and 27 to draw the full magentization curve. The clusters with $N=30$ and $33$ are used only to calculate the high field region of the full magnetization.  For $S=1$, we use the clusters with $N=12$, 15 and 18 to draw the full magentization curve. The cluster with $N=21$ is used at higher field region. The magnetization curves of the $S=1/2$ and $S=1$ cases are shown in Fig. \ref{uni}. Apparently, there exists a magnetization plateau with magnetization $M = M_{\scriptsize\rm s}/3$ where $M_{\scriptsize\rm s} (\equiv NS)$ is the saturation magnetization. The end points of the plateaux are plotted against the system size in Fig. \ref{pla}. It is evident that the size dependence of the width of the plateau is weak in the $S=1$ case while it is rather strong in the $S=1/2$ case. This would be due to the localized nature of the HSS state realized in $S=1$ KHAF.\cite{kh1} However, even for the $S=1/2$ case, the width does not strongly depend on the system size for $N \geq 21$.  Therefore we may expect that the 1/3-plateau survives in the thermodynamic limit both for $S=1/2$ and $S=1$.

Recently, Zhitomirsky\cite{zhito1} has pointed out that the magnetization plateau at $M_{\scriptsize\rm s}/3$ in the classical KHAF is stabilized by the order-from-disorder mechanism due to thermal fluctuation. Similar phenomenon is also pointed out in the classical  triangular\cite{miya} and frustrated square lattice Heisenberg antiferromagnet.\cite{zhito2} In these cases, the plateau state consists of  the collinear $\uparrow\uparrow\downarrow$ structure on the triangular plaquettes, which is selected by the large entropy gain due to the soft modes associated with this structure. It should be noted that the classical  $\uparrow\uparrow\downarrow$  states themselves are still infinitely degenerate due to infinite number of tiling patterns of $\uparrow\uparrow\downarrow$ structure on the triangular plaquettes.  It is known that the quantum fluctuation also plays a similar role in the  triangular lattice.\cite{cg1,jns1}. Therefore this mechanism should also work in quantum KHAF. In the quantum case, the ground state is stabilized because of the low zero point energy of the above mentioned soft spin wave modes. In analogy with the classical case, the plateau state in the quantum KHAF can be regarded as the quantum superposition of the $\uparrow\uparrow\downarrow$ tiling structures. 

 From this point of view, the absence of the plateau for the honeycomb lattice is easily understood. Although the honeycomb lattice, as well as the kagom\'e lattice, can be regarded as a depleted triangular lattice, all triangles are depleted in the honeycomb lattice so that no plateau remains.

\subsection{Distorted kagom\'e lattice}

There are various ways to introduce distortion in the present model. To be specific, we concentrate on the following three cases:  (A) $\JA=J(1+\delta), \JB=J, \JC=J(1-\delta)$ ,  (B) $\JA=J, \JB=J(1-\delta), \JC=J(1-\delta)$ and  (C) $\JA=J, \JB=J(1-\delta), \JC=0$. The numerical calculation is performed with $N=18$ and 27 for $S=1/2$ case and with $N=18$ for $S=1$ case.

\subsubsection{{\rm Case (A)}: $\JA=J(1+\delta), \JB=J, \JC=J(1-\delta)$}

\begin{figure}
\centerline{\epsfxsize=80mm\epsfbox{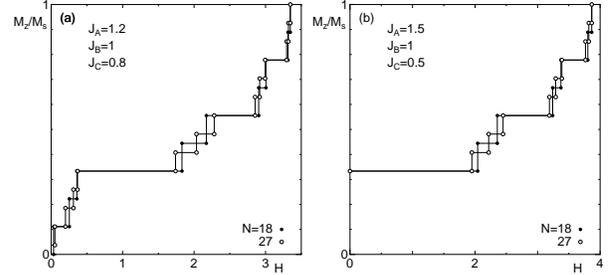}}
\caption{The magnetization curves of $S=1/2$ distorted KHAF with (a) $\delta=0.2$ and (b) 0.5 for $N = 18$ and 27 for case (A). }
\label{magdisth}
\end{figure}

\begin{figure} 

\centerline{\epsfxsize=80mm\epsfbox{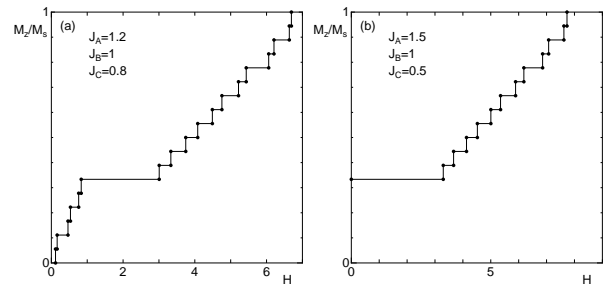}}
\caption{The magnetization curves of $S=1$ distorted KHAF with (a) $\delta=0.2$  and (b) 0.5 for $N = 18$  for case (A).}
\label{magdist1}
\end{figure}

The magnetization curves  are shown in Fig. \ref{magdisth} for $S=1/2$ and Fig. \ref{magdist1} for $S=1$, respectively. It is evident that the plateau region is enhanced as $\delta$ increases and eventually the ferrimagnetic state is realized for large enough $\delta$. 

The enhancement of the plateau and the appearance of the ferrimagnetic state for large $\delta$ is understood as follows. For large $\delta$, the 6 spins around each A-hexagon approximately form a 6-spin singlet state. The number of remaining spins is $N/3$.  The effective coupling between these spins are weak so that the magnetization of these spins easily saturate.  To magnetize further, large energy to break up the singlet on A-hexagon is required. Thus, the plateau is enhanced by distortion. Especially, the effective interaction between the 'alive' spins mediated by the second strongest bond B is ferromagnetic. Therefore, if $\JA \gg \JB \gg \JC$, the 'alive' spins form ferrimagnetic state with magnetization $M = M_{\scriptsize\rm s}/3$. 

 It should be remarked that this structure of the plateau state is consistent with the picture of the plateau state of the undistorted KHAF which is the quantum superposition of the $\uparrow\uparrow\downarrow$ tiling structures. In the plateau state of the strongly distorted KHAF, one spin remains 'alive' for each triangle and all of them are up-spins. The remaining two spins on the same tringle are coupled via $\JA$-bond and antiferromagnetically correlated. This is a kind of superposition state of $\uparrow\uparrow\downarrow$ states although the position of one of the up-spin is strongly limited. Therefore, the plateau state and the ferrimagnetic state in the distorted KHAF are continuously connected with that in the undistorted KHAF.

\subsubsection{{\rm Case (B)}: $\JA=J, \JB=J(1-\delta), \JC=J(1-\delta)$}
\begin{figure}
\centerline{\epsfxsize=80mm\epsfbox{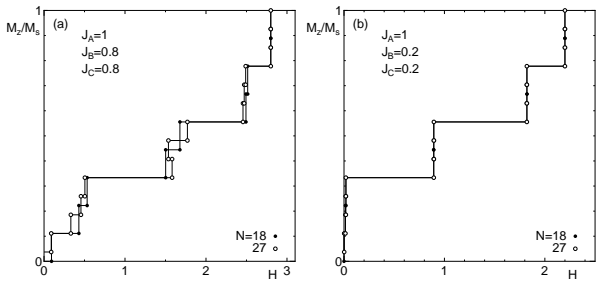}}
\caption{The magnetization curves of $S=1/2$ distorted KHAF with (a) $\delta=0.2$ and (b) 0.8  for $N = 18$ and 27 for case (B). }
\label{singh}
\end{figure}

\begin{figure} 
\centerline{\epsfxsize=80mm\epsfbox{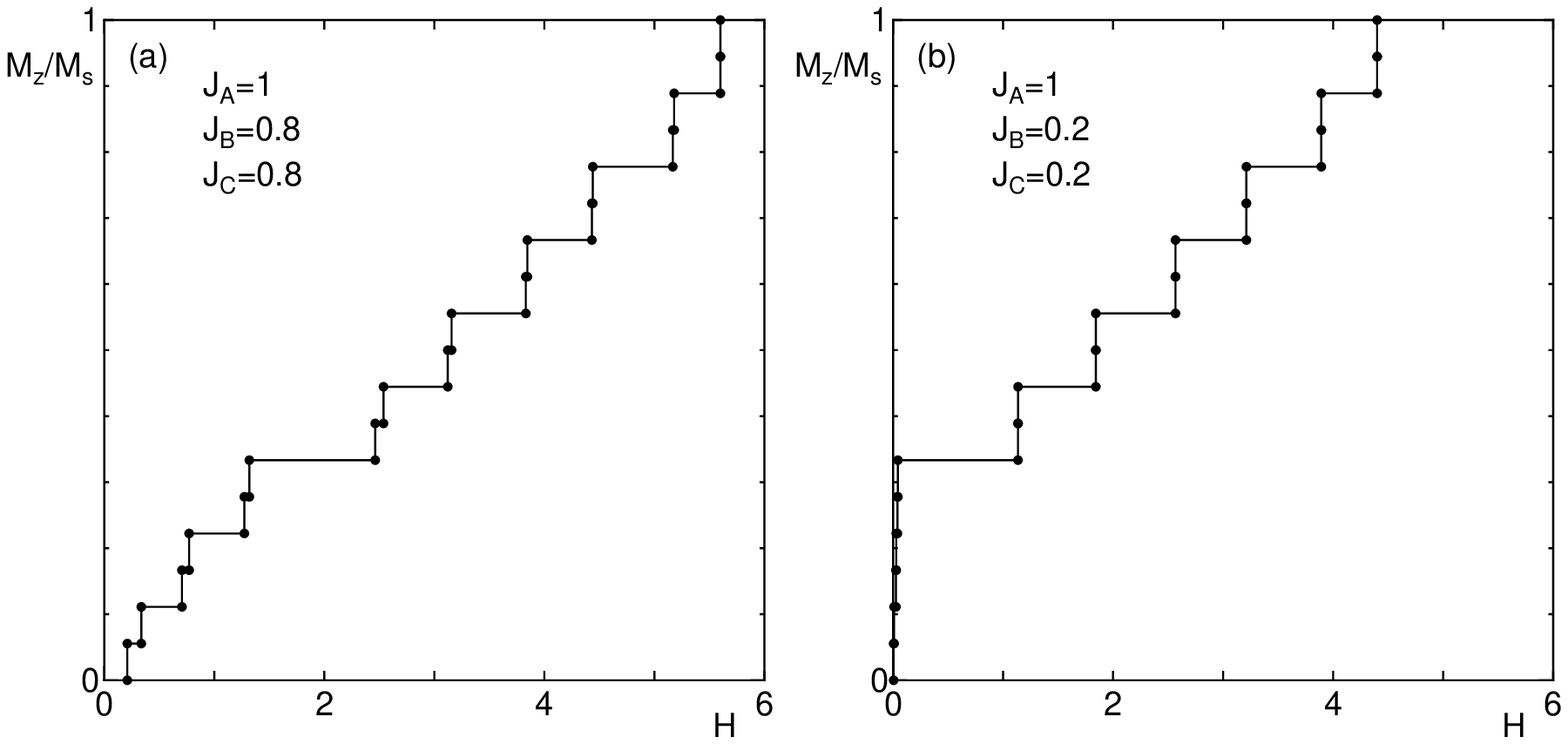}}
\caption{The magnetization curves of $S=1$ distorted KHAF with (a) $\delta=0.2$  and (b) 0.8  for $N = 18$ for case (B). }
\label{sing1}
\end{figure}

In this case, the ground state remains singlet as shown in Fig. \ref{singh} and Fig. \ref{sing1}. The physical picture of the ground state in this case is discussed in detail in  ref. \citen{khnew} which is summarized as follows. In the strong distortion limit $\JA \gg \JB=\JC$, the strongest effective interaction among 'alive' spins is the next nearest neighbour antiferromagnetic interaction according to the perturbational calculation in ref. \citen{khnew}. If other weaker effective interactions are neglected, the  whole lattice of 'alive' spins are decomposed into three equivalent sublattices of kagom\'e type. These 'alive' spins again form singlet states on large hexagons which are three times larger than the original kagom\'e lattice and the ground state of the whole system is again singlet.\cite{khnew} 

 The slight backbending of the magnetization curve is observed above the plateau for the $N=27$ cluster with $S=1/2, \delta=0.2$ (Fig. \ref{singh}(a)). This implies the metamagnetic behavior, if it remains in the thermodynamic limit. This phenomenon, however, takes place only in a narrow paramater range around $\delta=0.2$ of the $N=27$ cluster. Within the present calculation, therefore, it is not clear if this is the artifact of the finite size effect or truely thermodynamic effect. This problem is left for future studies.


\subsubsection{{\rm Case (C)}:$\JA=J, \JB=J(1-\delta), \JC=0$}

\begin{figure}
\centerline{\epsfxsize=80mm\epsfbox{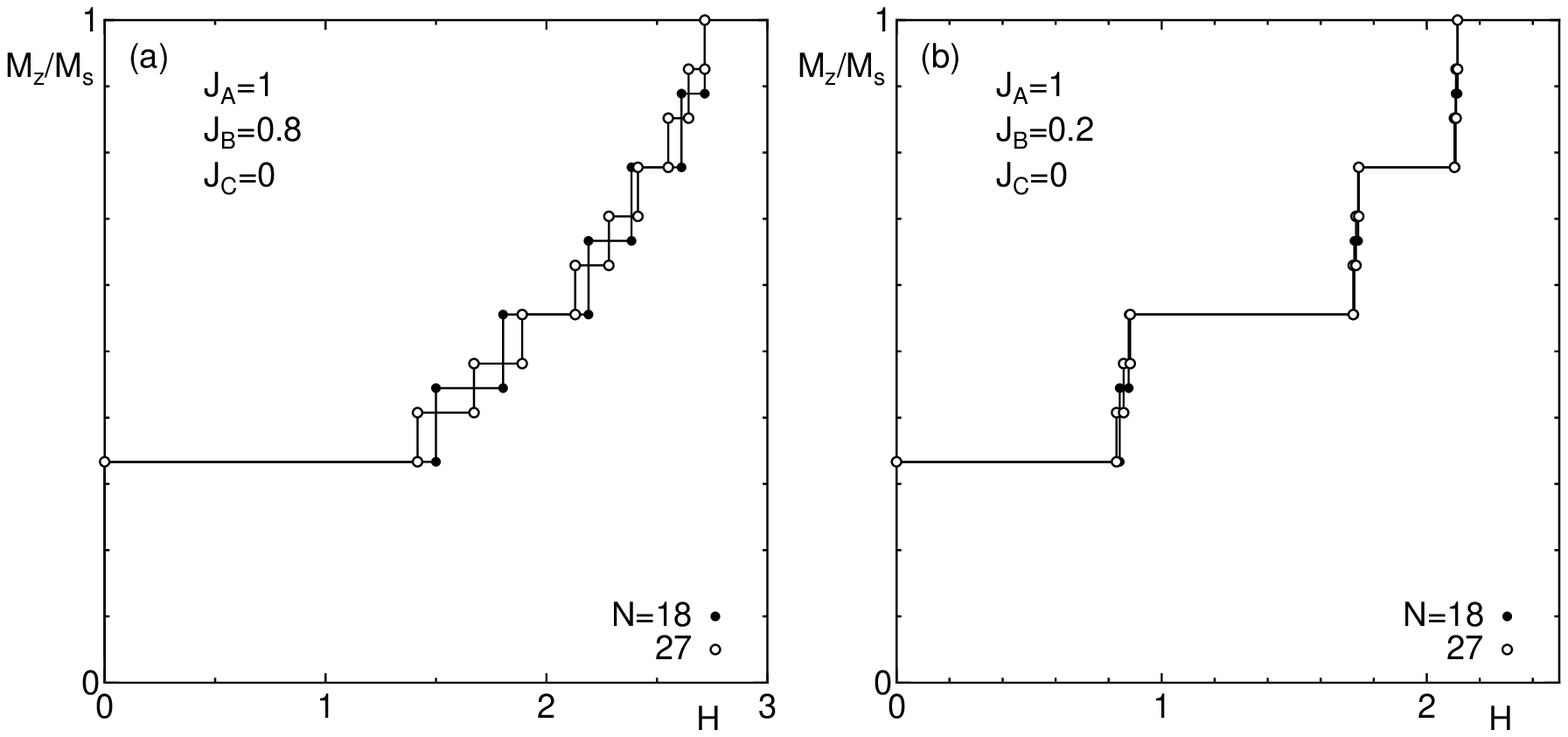}}
\caption{The magnetization curves of $S=1/2$ distorted KHAF with (a) $\delta=0.2$ and (b) 0.8  for $N = 18$ and 27  for case (C). }
\label{ferrih}
\end{figure}
\begin{figure}
\centerline{\epsfxsize=80mm\epsfbox{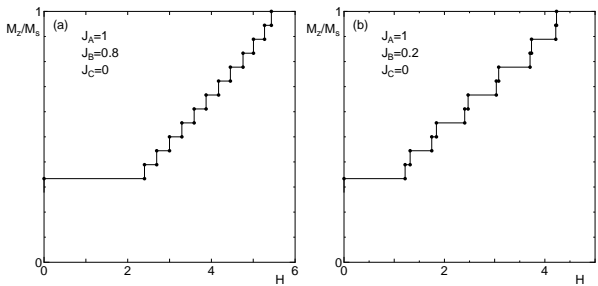}}
\caption{The magnetization curves of $S=1$ distorted KHAF with (a) $\delta=0.2$  and (b) 0.8 for $N = 18$ for case (C). }
\label{ferri1}
\end{figure}
In contrast to case (B), the ground state is always ferromagnetic as shown in Fig. \ref{ferrih} and  Fig. \ref{ferri1}. This is easily understood by the Lieb-Mattis theorem.\cite{lm1} For $\JC=0$, the whole lattice can be decomposed into two sublattices. One of the sublattice has $2N/3$ sites and the other sublattice, which is coupled antiferromagnetically with the former, has $N/3$ sites. As a result, the ground state has total magnetization $NS/3$. This is also consistent with the argument in the strong distortion limit at the end of the last paragraph of case (A), because the effective interaction between the 'alive' spins are all ferrimagnetic for $\JC=0$. 

For distorted cases, there are minor plateaux at $M_z=(1/3+2m/9)M_{\scriptsize\rm s}, m=1,2$ for $S=1/2$  and at $M_z=(1/3+m/9)M_{\scriptsize\rm s}, m=1,..,5$ for $S=1$ above the main plateau at $M_z=M_{\scriptsize\rm s}/3$, although it is not clear if they remain in the thermodynamic limit.  Of course, it is obvious that these plateaux appear for  $\JB, \JC \ll \JA$, because these plateaux corresponds to the magnetization steps of isolated hexagons. Actually, in the undistorted case, such plateaux do not exist. Surprisingly, however, they also appear even if $\JB$ and $\JC$ are close to $\JA$ as far as $\JB =\JC$. This is understood in the following way. If $\JB = \JC$, the effective field due to the polarization of 'alive' spins are uniform on the six sites around the A-hexagon, so that it only cause the shift of effective magnetic field and the positions of the plateaux are essentially those for the isolated hexagons. If the difference between $\JB$ and $\JC$ is large, the effective field is nonuniform and the 6-spin eigenstates with different magnetization around the A-hexagon are strongly mixed up and the plateaux are smeared. In case (A), it should be noted that such plateaux are smeared for $S=1$ but not for $S=1/2$ for $\delta=0.2$. Presumably, this is due to the stronger quantum effect in the $S=1/2$ case. In the $S=1/2$ case, the magnetization steps for the isolated hexagon is strongly quantized compared to the $S=1$ case so that the steps for the $S=1/2$ case are not easily smeared out. If the distortion is weaker, however, the minor plateaux are smeared out even in the $S=1/2$ case. Therefore this difference is not an essential difference between $S=1/2$ and $S=1$ cases but is rather a quantitative one.

\section{Summary and Discussion}
The magnetization process of the spin-1/2 and 1 KHAF is studied by means of the exact diagonalization method. The magnetization plateau at 1/3 of the full magnetization is found. The physical picture of the plateau state is explained as a quantum superposition of the $\uparrow\uparrow\downarrow$ tiling states. It is shown that this plateau is enhanced by the $\sqrt{3}\times\sqrt{3}$ lattice distortion and the ferrimagnetic ground state is realized for large enough distortion in cases (A) and (C). The physical origin of the enhancement of the plateau and appearance of the ferrimagnetic state is explained from the argument in the strong coupling regime. It is also argued that the plateau state in the strong distortion limit is continuously connected with that of the uniform case. The interpretation of the bahavior of the minor plateaux in the high field regime is also given.

It should be interesting to observe  the magnetization plateau experimentally. For \mpynn, the effective exchange coupling between the $S=1$ spins is given by $J \simeq 0.78K $ so that $J/g\mu_{\scriptsize\rm B} \simeq 0.58T$.\cite{wada1,awaga1,wata1}  Thus the magnetic field at the plateau is easily feasible. The width of the plateau is around $J$. Therefore, the plateau would be clearly observed if the magnetization measurement at temperatures much below 1K is carried out. 

The author thanks M. E. Zhitomirsky for attracting the author's attention to ref. \citen{zhito1,zhito2} and for useful comments to the earlier version of this work. The numerical calculation is performed using the HITAC SR8000 at the Supercomputer Center, Institute for Solid State Physics, the University of Tokyo and the HITAC SR8000 at the Information Processing Center, Saitama University.  The diagonalization program is based on the TITPACK ver.2 coded by H. Nishimori and KOBEPACK/1 coded by T. Tonegawa, M. Kaburagi and T. Nishino. This work is supported by the Grant-in-Aid for Scientific Research from the Ministry of Education, Science, Sports and Culture.

\end{document}